\documentclass[preprint,12pt]{elsarticle}
\usepackage[cp1251]{inputenc}
\usepackage[english, russian]{babel}
\usepackage{hyphenat}
\usepackage{tabularx}
\usepackage{amsmath}
\usepackage{subfigure}
\usepackage [format=hang, margin=10pt,labelfont=bf,justification=centerlast, singlelinecheck=true]{caption}

\usepackage{amssymb}

\begin{document}

\begin{frontmatter}



\title{KINETICS OF IONS DURING THE DEVELOPMENT OF PARAMETRIC INSTABILITY OF INTENSIVE LANGMUIR WAVES IN PLASMA}


\author{A.V. Kirichok}
\ead{sandyrcs@gmail.com} 
\author{V.M. Kuklin}
\author{A.V. Pryjmak} 
\address{Kharkov National University, Institute for High Technologies, \\
4 Svobody Sq., Kharkov 61022, Ukraine}

\author{A.G. Zagorodny} 
\address{Bogolyubov Institute for Theoretical Physics, \\
14-b, Metrolohichna str., Kiev, 03680, Ukraine}

\selectlanguage{english}

\begin{abstract}
Nonlinear regimes of one-dimensional parametric instabilities of long-wave plasma waves are considered for the cases when the average plasma field energy density is less (Zakharov's model) or greater (Silin's model) than the plasma thermal energy density. The hybrid models presented in this paper treat the electrons as a fluid by way of an equation for the  high-frequency wave and treat the ions kinetically with a super-particle simulation. This makes possible consideration of non-resonant particles trapped by high-frequency oscillations and the influence of trajectory crossing on the development of the parametric instability. The evolution of ion energy distribution is studied. It is shown that after saturation of the instability, the ion kinetic energy density normalized to the initial field energy density is of the order of the ratio of linear growth rate to the plasma frequency, for the case when the initial field energy far exceeds the plasma thermal energy. In this case, the ion energy distribution is different from the Maxwellian.In the opposite case of hot plasma, the ions acquire a part of initial field energy, which is approximately equal to the half of ratio of initial Langmuir field energy to the plasma thermal energy. At this, the ion kinetic energy distribution is close to the Maxwellian, and it is reasonably to speak about ion temperature. The crossing of ion trajectories in the surrounding of density cavities is a reason of instability quenching in both cases.

\end{abstract}


\end{frontmatter}



\section{Introduction}

\selectlanguage{english}

The interest in parametric instability of intensive Langmuir waves, which can be easily excited in the plasma by various sources \cite{Silin.1961}, \cite{Basov.1964}, \cite{Dawson.1964}, \cite{Pashinin.1971}, \cite{Buts.2006}, \cite{Fainberg.2000}, \cite{Kuzelev.1990}, \cite{Shapiro.1976}, \cite{Kondratenko.1988} was stipulated, in particular, by the new possibilities in heating of electrons and ions in plasma. The correct methods for description of parametric instability of long-wave plasma waves were developed in the pioneering works of V.P. Silin \cite{Silin.1965} and V.E. Zakharov \cite{Zakharov.1967}, \cite{Zakharov1.1967}, \cite{Zakharov.1972}. The theoretical concepts of \cite{Silin.1965} were confirmed by the early numerical experiments on the one-dimensional modelling of parametric decay of plasma oscillations \cite{Kruer.1970} (see also \cite{Aliev.1965},  and review \cite{Silin.1973}). However, the greatest interest has been expressed by experimenters in the mechanism of dissipation of wave energy discovered by V.E. Zakharov. The analytical studies, laboratory-based experiments and numerical simulations, performed at an early stage of studying these phenomena \cite{Kruer.1988}, \cite{Ivanov.1974}, \cite{Kim.1974}, have confirmed the fact that in some cases a significant part of the pump field energy turns during the instability development into the energy of short-wave Langmuir oscillations attended with bursts of fast particles \cite{Andreev.1977}, \cite{Kovrizhnykh.1977}, \cite{Antipov.1976}, \cite{Sagdeev.1980}, \cite{Wong.1984}, \cite{Cheung.1985}, \cite{Zakharov.1989}. 

In this paper, we compare the models of Silin and Zakharov by the example of one-dimensional description. The choice of one-dimensional approach, as was noted by J. Dawson \cite{Dawson.1962}, <<… often keeps the main features of the processes, but simplifies their description and leads to a fuller understanding of what the important phenomena are>>. Of particular interest is the process of ion heating, so we use in this work the super-particle (or finite-sized particle) description for ions because the account of inertial effects can be significant just at the nonlinear stage of the process \cite{Kuklin.1990}.

It was observed in \cite{Kuklin.1990}, \cite{Clark.1992} that simulation with using of the so-called hybrid model (incorporating one of the Zakharov equations for the high-frequency waves and using a particle simulation of the ions) demonstrates that fluctuations of ion density are rather significant and accelerate the development of parametric instability. The non-resonant interaction between super-particles-ions and high-frequency plasma oscillations, along with the trapping of ions into the potential wells, formed by these oscillations, leads to instability of the density cavities resulting from the modulation instability.

In paper \cite{Clark.1992} the hybrid model was compared with Zakharov’s hydrodynamic model. Due to higher level of ion density fluctuation, the number of cavities in the hybrid model appears to be significantly greater than in the Zaharov model and their depth is less. Integral characteristics of both models are essentially identical. Note that both the hydrodynamic description within framework of the Zaharov model \cite{Clark.1992} and description based on the kinetic equations \cite{Henri.2011}, in which non-resonant interactions such as “particle -- finite amplitude wave” and the trapping of particles by the wave are ignored. As a sequence, the resulting cavities remain stable until the moment when the high frequency plasma field is “burned out” due to the Landau damping -- the process that can be better described with using the method of finite-sized particles, as was reasonably pointed out in \cite{Clark.1992}.

In the simulation of one-dimensional ion dynamics below, we have used $2\cdot 10^{4} $ super-particles, which is equivalent to the number of ions about $(2 \cdot 10^{4} )^{3} \sim 10^{13} $ for three-dimensional case that is in agreement with the conditions of most experiments. Thus, the interaction between modeling super-particles and plasma oscillations in this simulation is quite good corresponds to the interaction for interaction between real particles and plasma waves of course with regard to the inherent limitations of one-dimensional description. Nevertheless, there is reason to believe that the transfer of field energy to ions within framework of the hybrid model corresponds to the real conditions of ion heating by intense Langmuir oscillations in plasma.  



\section{The hybrid models of parametric instability}

  \subsection{The hybryd model based on the Silin equations, $|E_{0} |^{2} /4 \pi \gg n_{0} T_{e}$} 
  When the intensity of external electric field is much greater than the specific thermal energy of plasma electrons $W=|E_{0} |^{2} /4\pi \gg n_{0} T_{e} $, it is reasonable to explore the approach presented by V.P. Silin \cite{Silin.1973}. 
  
  Let consider a one-dimensional plasma system, where an intense plasma wave with the wavelength $\lambda _{0} $ and frequency $\omega _{0} $ is excited by an external source. This intense wave will be referred to as the pumping wave.  Since the parametric instability results in the growth of oscillations with rather small wavelength $\lambda <<\lambda _{0} $, the pumping wave can be considered as spatially uniform within the region of interaction:

\begin{equation}
E_{0} =-i(|E_{0} |\exp \{ i\omega _{0} t+i\phi \} -|E_{0} |\exp \{ -i\omega _{0} t-i\phi \} )/2,             
\label{eq1}
\end{equation}

\noindent where $|E_{0} |$and $\phi $ are the slowly varying wave amplitude and phase correspondingly, $\omega_{0} $ is external wave frequency, $n_{0}$ and $T_{e}$ are the density and temperature of plasma electrons. Charged particles of plasma oscillate under the action of the electric field and their velocities can be written as $u_{0\alpha } =-\left({e_{\alpha } \left|E_{0} \right| / m_{\alpha } } \omega_{0}\right) \cos \phi =-\omega _{0} b_0 \cos \phi $, where $b_0={e_{\alpha } \left|E_{0} \right| /m_{\alpha }\omega_0^2 }$ is the particle oscillation amplitude. 

The equations, governing the nonlinear dynamics of the parametric instability of intensive plasma wave were derived in \cite{Kuklin.2013}. The equations for high-frequency plasma field spectrum modes $E=\sum _{n}E_{n}  (t)\cdot \exp (ink_{0} x)$ (plasma electrons are considered as fluid and described by hydrodynamic equations) take a form

\begin{multline}
\frac{\partial E_{n} }{\partial t} -i\frac{\omega _{pe}^{2} -\omega _{0} ^{2} }{2\omega _{0} } E_{n} -\frac{4\pi \omega _{pe} \nu _{i,n} }{k_{0} n} J_{1} (a_{n} ) \exp (i\phi )- \\
-i\frac{\omega _{0} }{2en_{0} } \sum _{m}\nu _{i,n-m}   [E_{-m}^{*} J_{2} (a_{n-m} )e^{2i\phi } +E_{m}  J_{0} (a_{n-m} )]=0.           
\label{eq2}
\end{multline}
\noindent Here $\omega _{pe} =\sqrt{4\pi e^{2} n_{0} /m_{e} } $ is the background electron plasma frequency, $e$ and $m_{e} $ are the mass and the magnitude of the charge of an electron, $M$is the mass of an ion, $E_{n} =|E_{n} |\cdot \exp (i\psi _{n} )$is a slowly varying complex amplitude of the electric field of electron plasma oscillations, which wavenumber is $k_{n} =nk_{0} $, $k_{0} ={2\pi /L}$, where $L$ is a characteristic dimension of the plasma system, $\nu _{i} =\sum _{n}\nu _{i,n}  (t)\cdot \exp (ink_{0} x)$ is the ion charge density, $a_{n} =a\cdot n$, $n,m$ are integers which are not equal to zero and $\pm 1$, e.g. ${a_n} = n{k_0}b = n\left( e{k_0}{E_0}/{m_e} \omega _0^2\right)$.

The motion equations for ion super-particles can be written as follows 

\begin{equation}
\frac{d^{2} x_{s} }{dt^{2} } =\frac{e}{M} \sum _{n} \bar{E}_{n}  \exp \{ ik_{0} nx_{s} \},               
\label{eq3}
\end{equation}

\noindent and the ion density can be determined from

\begin{equation}
n_{i,n} =\nu _{i,n} /e= \frac{n_{0} k_{0} }{2\pi } \int _{-\pi /k_{0} }^{\pi /k_{0} }\exp [-ink_{0}  x_{s} (x_{0} ,t)] dx_{s0}. 
\label{eq4}
\end{equation}

The slowly varying electric field strength $\bar{E}_{n} $, acting on the ions, is equal 

\begin{multline}
\bar{E}_{n} =-\frac{4\pi i}{k_{0} n} \nu _{i,n} [1-J_{0}^{2} (a_{n} )+\frac{2}{3} J_{2} ^{2} (a_{n} )]+\\
+\frac{1}{2} J_{1} (a_{n} )[E_{n}  e^{-i\phi } -E_{-n}^{*}  e^{i\phi } ]- 
-\frac{ink_{0} }{16\pi en_{0} } J_{0} (a_{n} )\sum _{m}E_{n-m}  E_{-m}^{*}  -\\
-\frac{ik_{0} J_{2} (a_{n} )}{16\pi en_{0} } \sum _{m}(n-m)[E_{n-m}  E_{m}  e^{-2i\phi } +E_{m-n}^{*}  E_{-m}^{*}   e^{2i\phi } ], 
\label{eq5}
\end{multline}

The equation for uniform component of the electric field $E_{0} =\left|E_{0} \right|\exp (i\phi )$ can be written as

\begin{equation}
\frac{\partial E_{0} }{\partial t} =-\frac{\omega _{0} }{2en_{0} } \sum _{m}\nu _{i,-m}   [E_{-m}^{*}  J_{2} (a_{m} )e^{2i\phi } +E_{m}  J_{0} (a_{m} )].
\label{eq6} 
\end{equation}

Note, that the values with subscripts of different signs are independent. In Eqs.(\ref{eq2})-(\ref{eq6}), we have used the formula \cite{Dwight.l96l}

\begin{equation}
\exp \{ ia \sin\Phi \} =\sum _{m=-\infty }^{\infty }J_{m}  (a) \exp \{ im\Phi \},
\label{eq7}
\end{equation}

\noindent where $J_{m}(x)$ is the Bessel function.

The normalized frequency shift $\Delta =(\omega _{pe}^{2} -\omega _{0} ^{2} )/2\delta \omega _{pe} $ reaches the value of $({m_{e}/2M} )^{1/3} J_{1} ^{2/3} (a_{n} )$ for a mode with the maximal growth rate of parametric instability $\delta $ \cite{Silin.1973}

\begin{equation}
\delta /\omega _{pe} =\frac{i}{\sqrt[{3}]{2} } \left(\frac{m_{e} }{M} \right)^{1/3} J_{1} ^{2/3} (a_{n} ).
\label{eq8}
\end{equation}

\subsection{The hybryd model based on the Zakharov equations, $|E_{0} |^{2} /4\pi \ll n_{0} T_{e}$} 
As shown in \cite{Kuklin.2013}, Eqs. (\ref{eq2})-(\ref{eq6}) are coincident with equations, obtained in \cite{Kuznetsov.1976} after following substitutions: $(\omega _{pe}^{2} -\omega _{0} ^{2} )/2\omega _{0} \to (\omega _{pe}^{2} -\omega _{0} ^{2} +k^{2} _{0} n^{2} v_{Te}^{2} )/2\omega _{0} $ and $E_{0} \to -iE_{0} $, $E_{0}^{*} \to iE_{0}^{*} $ under condition $a_{n} \ll 1$, that means that $J_{1} (a_{n} )\approx a_{n} /2$, $J_{0} (a_{n} )\approx 1$, $J_{2} (a_{n} )\approx a_{n}^{2} /8$

\begin{equation}
\frac{\partial E_{n} }{\partial t} -i\frac{\omega _{pe}^{2} -\omega _{0} ^{2} +k^{2} _{0} n^{2} v^{2} _{Te} }{2\omega _{0} } E_{n} -i\frac{\omega _{0} }{2n_{0} }  \left( n_{i,n} E_{0} +\sum _{m\ne 0}n_{i,n-m} E_{m}  \right) =0.
\label{eq9}
\end{equation}

The slowly varying electric field amplitude in this case takes the form 

\begin{equation}
\bar{E}_{n} =-\frac{ik_{0} ne}{4m\omega _{p} ^{2} } \left(E_{n} E_{0}^{*} +E_{0} E_{-n}^{*} +\sum _{m\ne 0,n}E_{n-m} E_{-m}^{*}  \right),
\label{eq10}
\end{equation}

\noindent that enables description of ions using the super-particle method with use of Eqs.(\ref{eq3})-(\ref{eq4}). The pump wave amplitude $E_{0}$ is governed by equation

\begin{equation}
\frac{\partial E_{0} }{\partial t} -i\frac{\omega _{0} }{2n_{0} }  \sum _{m}n_{i,-m} E_{m}  =0.
\label{eq11}
\end{equation}

In this case, the growth rate of the parametric instability normalized to the plasma frequency is \cite{Kuznetsov.1976}

\begin{equation}
\delta /\omega _{pe} =\left( \frac{|E_{0} |^{2} }{8\pi n_{0} T_{e} } \frac{m_{e} }{M} \right)^{1/2} =\left(\frac{W}{2 n_{0} T_{e} } \frac{m_{e} }{M} \right)^{1/2}.
\label{eq12}
\end{equation}

\section{The simulation parameters}

The purpose of this paper is to clarify the characteristics of the dynamics of modulation instability both for the cases of non-isothermal hot and cold plasma within framework of the hybrid models.
Each model is considered for the two cases of light and heavy ions. The parameters of simulation are presented in Table 1.

\begin{table}[b]  \footnotesize
\caption{Simulation parameters for the hybrid models}
\begin{tabularx}{\textwidth}{lXX} \hline \hline 
Model & Light ions

${M}/{m_{e}}  =2\cdot 10^{3}$
& Heavy ions 

 $m_{e}/M= 8\cdot 10^{-6} $
 \\ \hline \noalign{\smallskip} 
 Silin   &  $(m_e/M) (\omega ^{2} _{p} /\delta ^{2})=0.43$
 
 ${\delta /}{\omega _{0} } =0.44 \cdot ({m_{e} }/{M} )^{1/3} =0.034$ 
 
 ${\omega _{0} }/{\delta } \approx {\omega _{pe} }/{\delta } =29.4$  & $(m_e/M) (\omega ^{2} _{p} /\delta ^{2})=0.1$ 
 
 $\delta /{\omega _{0} } =0.44\cdot ({m_{e} }/{M} )^{1/3} =0.0088$ 
 
 $\omega _{0} /\delta  \approx {\omega _{pe} }/{\delta } =113.6$  \\ \hline \noalign{\smallskip}
Zakharov & $(m_e/M) (\omega ^{2} _{p} /\delta ^{2})=2{n_{0} T_{e} }/{W} =2 \cdot 10$

 $\omega _{0}/\delta  =2\left(\frac{n_{0} T_{e} }{W} \right)^{1/2} (\frac{M}{m_{e} } )^{1/2} =282.6$
 
  ${\delta }/{\omega _{0} } ={\delta }/{\omega _{pe} } = 3.5 \cdot 10^{-3} $ & 
  $(m_e/M) (\omega ^{2} _{p} /\delta ^{2})=2{n_{0} T_{e} }/{W} =2 \cdot 10$
  
   ${\omega _{0} }/{\delta } =2(\frac{n_{0} T_{e} }{W} )^{1/2} (\frac{M}{m_{e} } )^{1/2} =2234.4$ 
   
   ${\delta }/{\omega _{0} }={\delta }/{\omega _{pe} } =4.5\cdot 10^{-4} $ \\ 
   \hline 
\end{tabularx}
\label{Tab1}
\end{table}

Below, we have used the following initial conditions and parameters unless otherwise specified in the text. The number of super-particles, simulating the dynamics of ions is $0<s\le S=20000$. The super-particles are distributed uniformly over the interval $-1/2<\xi <1/2$, $\xi =k_{0} x/2\pi$, initial velocities of the super-particles are defined as $d\xi _{s} /d\tau |_{\tau =0} =v_{s} |_{\tau =0} =0$, the number of spectrum modes is $-N<n<N$, $N=S/100$. The initial normalized amplitude of the pumping wave is $a_{0} (0)=ek_{0} E_{0} (0)/m_{e} \omega _{pe}^{2} =0.06$,. The initial amplitudes of high-frequency plasma oscillations are defined by expression $e_{n} |_{\tau =0} =e_{n0} =(2+g_n) \cdot 10^{-3} $ for the Silin model and by expression $e_{n} |_{\tau =0} =e_{n0} =(0.5+g_n)\cdot 10^{-4} $ for the Zakharov model, where $g_n\in \left[0;1\right]$ is a random value, $ek_{0} E_{n} /m_{e} \omega _{pe}^{2} =e_{n}  \exp (i\psi _{n} )$. The initial phases of spectral modes $\psi _{n} |_{\tau =0} $ are also randomly distributed in the interval $0\div 2\pi $. For ion density fluctuations $n_{ni} $ and slowly varying electric field $\bar{E}_{n} $ we have used the following dimensionless representations: 
$$M_{n} =M_{nr} +iM_{ni} =n_{ni} \omega _{pe} /n_{0} \delta =\left(\omega _{pe}/{\delta } \right) \int _{-\pi /k_{0} }^{\pi /k_{0} }\exp (2\pi n \xi _{s} ) d\xi _{s0}  $$ 
\noindent and  
$$ek_{0} \bar{E}_{n} /m_{e} \omega _{pe}^{2} =E_{nr} +iE_{ni}.$$

The program, which implements a mathematical model of the problem under consideration, was developed with the use of JCUDA technology. JCUDA technology provides interface between CUDA (Compute Unified Device Architecture) and Java application. CUDA is a parallel computing platform and programming model created by NVIDIA. CUDA enables scientists to utilize the extreme computational power available on modern GPUs.

\begin{figure}[p]\center
\subfigure[Silin model]{\includegraphics[width=0.47\textwidth]{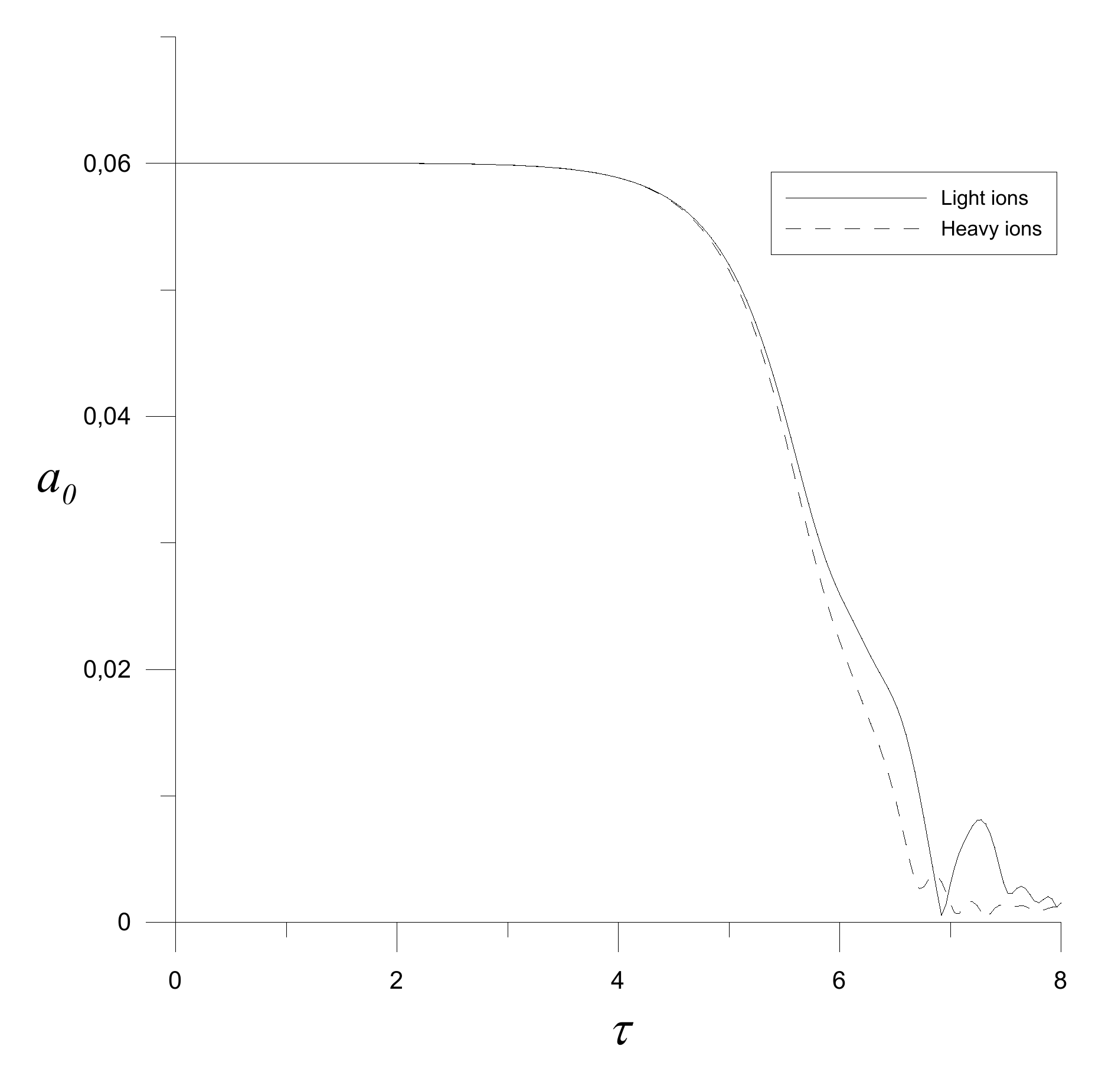}}  \hfill
\subfigure[Zakharov model]{\includegraphics[width=0.47\textwidth]{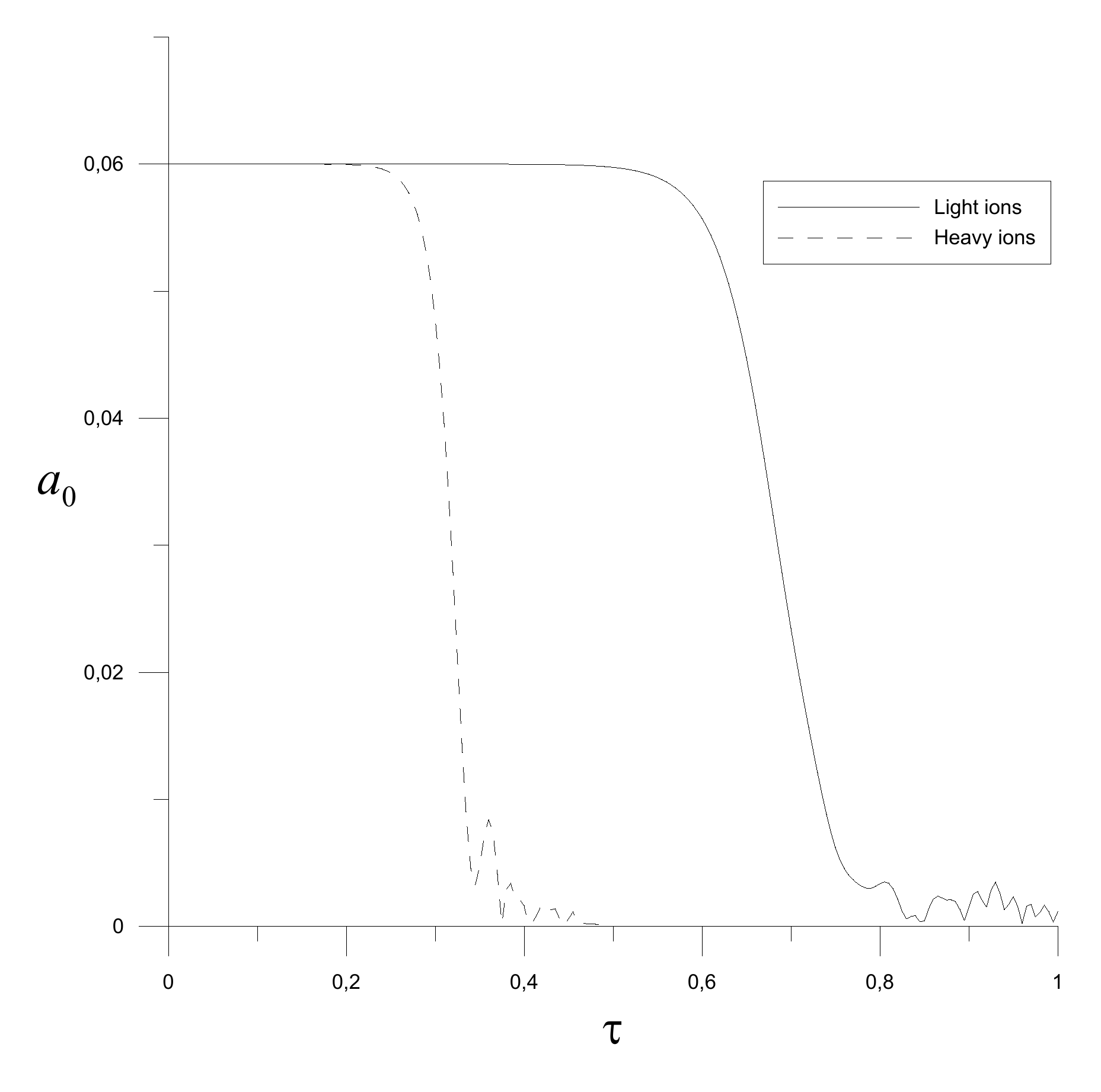}} \\

\caption{Evolution of the pump wave amplitude $a_{0}$.}
 \label{fig1}
\end{figure}

\section{Simulation results}

With development of the parametric instability, the ions trapped in the potential wells of cavities acquire the kinetic energy. At the nonlinear stage of the instability, the ion trajectories cross each other, the ion density perturbations are smoothed out and their characteristic scale increases. The relation between ion density perturbations and RF field becomes weaker and the instability saturates. The amplitude of the pump wave flattens out at rather low level after several oscillations (see Fig.\ref{fig1}).

\begin{figure}[p]\center
\subfigure[Silin model]{\includegraphics[width=0.47\textwidth]{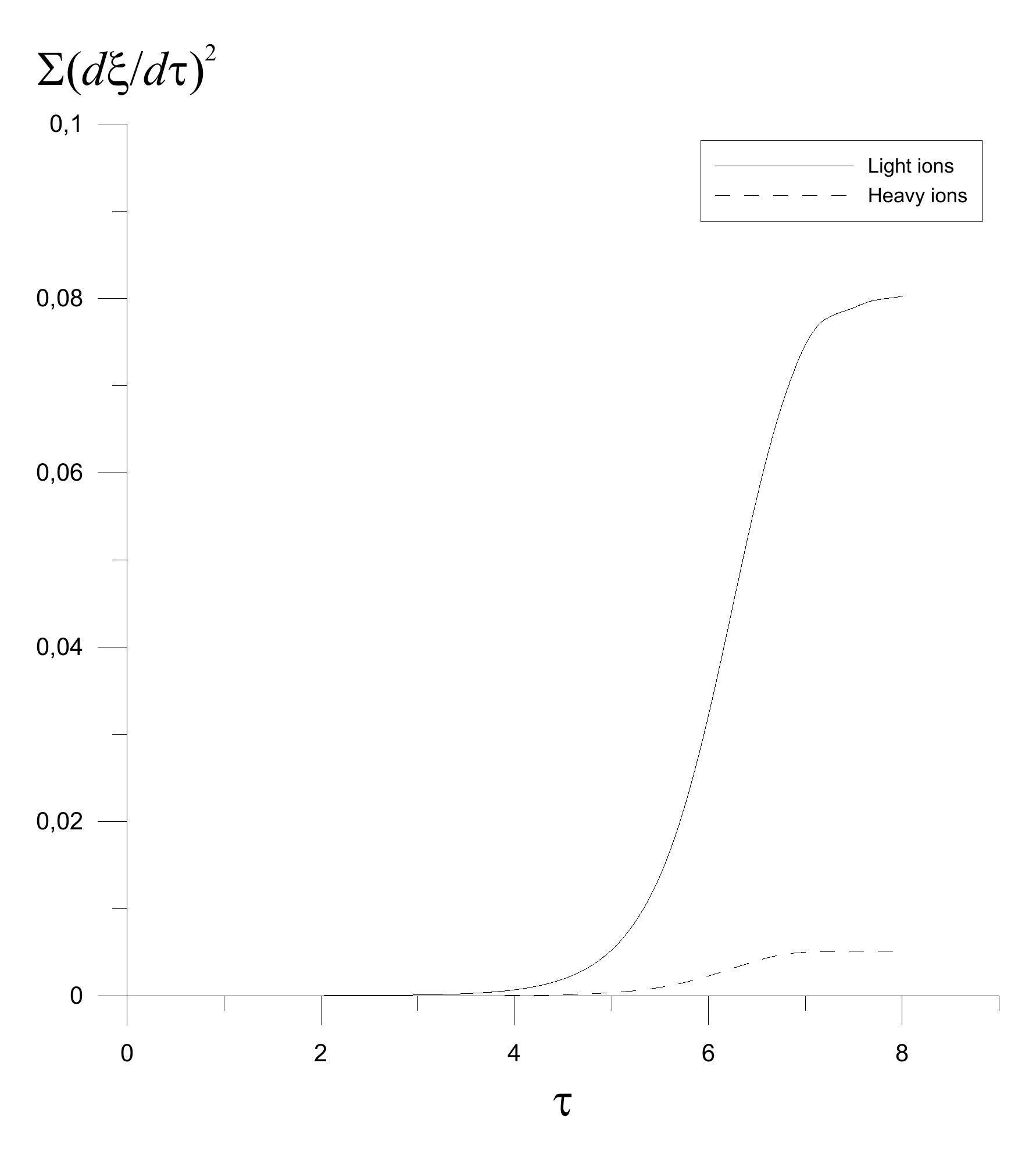}} \hfill
\subfigure[Zakharov model]{\includegraphics[width=0.47\textwidth]{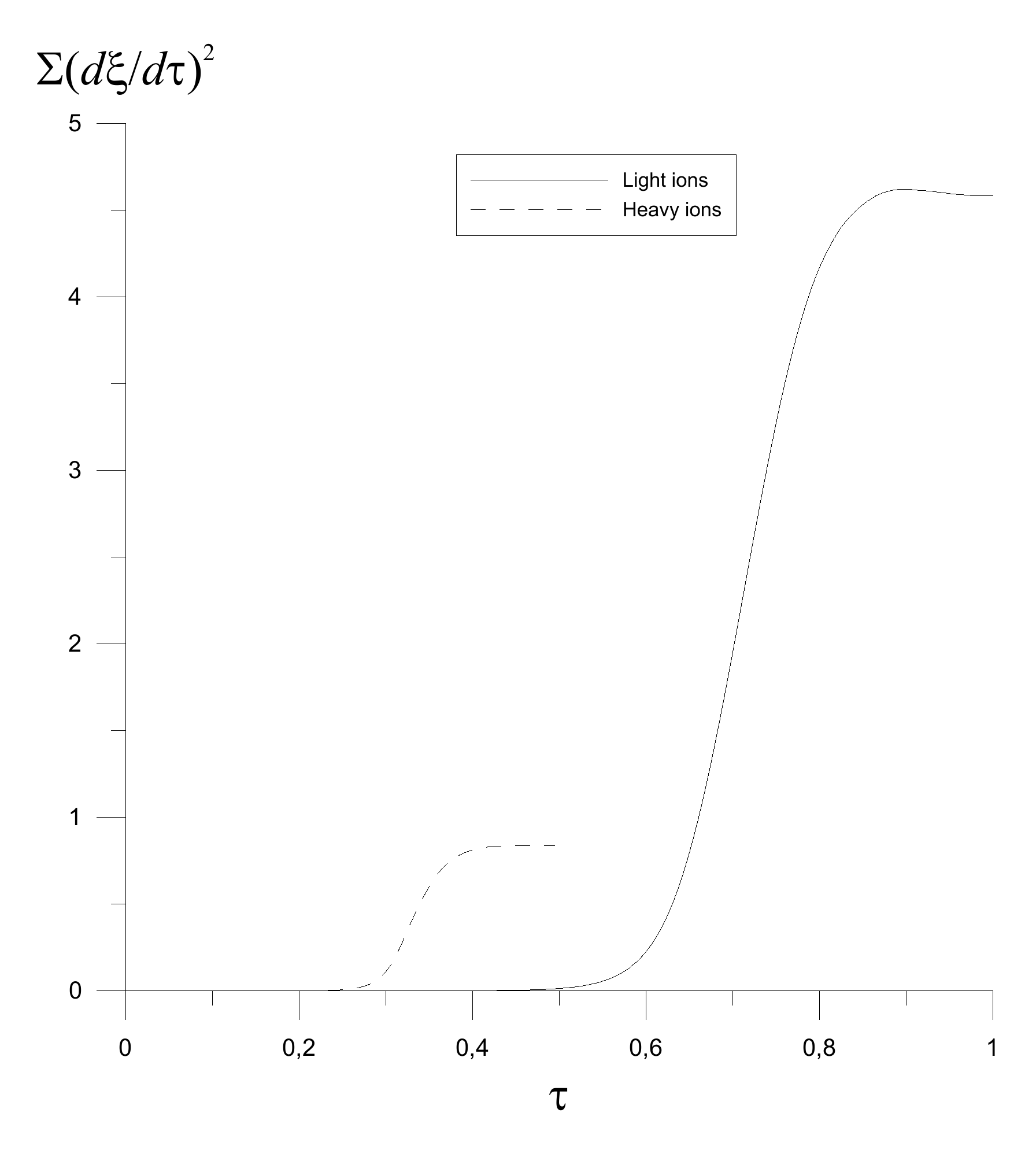}} 

\caption{Evolution of the quadratic sum  $\sum _{s}(d\xi _{s} /dt)^{2}  $, which is proportional to the ion kinetic energy.}

 \label{fig2}
\end{figure}

The main part of initial energy transfers from the pump wave to the short-wave Langmuir spectrum band.
The minor part of the initial energy converts to the kinetic energy of ions  (see Fig.\ref{fig2}).

The total kinetic energy of the ions located on the wave-length of the pump wave can be expressed through the sum of squared dimensionless velocities $I_{s} =\sum _{s}({d\xi _{s} }/{d\tau } ) ^{2} $ and the total number of super-particles $S$:
\begin{equation} \label{eq13} 
\frac{2\pi }{k_{0} } \left[\frac{1}{2} n_{0} M\left\langle \left({dx_{s} }/{dt} \right)^{2} \right\rangle \right]=I \frac{4\pi ^{2} \delta ^{2} Mn_{0} }{2k_{0} ^{2} S} \frac{2\pi }{k_{0} },  
\end{equation} 
where $\langle ({dx_{s} }/{dt} )^{2} \rangle $ is the ensemble average. The ratio of ion kinetic energy to the initial energy of intense long-wavelength Langmuir wave can be written as follows:

\begin{equation} \label{eq14} 
\frac{{\rm E} _{kin} }{W_{0} } =\frac{2\pi }{k_{0} } \left[\frac{1}{2} n_{0} M\left\langle \left({dx_{s} }/{dt} \right)^{2} \right\rangle \right]\left/{ \left[\frac{|E_{0} |^{2}} {4\pi}  \frac{2\pi }{k_{0} } \right]}\right. =2\pi ^{2} \frac{I}{a_{0} ^{2} S} \frac{M}{m}  \frac{\delta ^{2}}{\omega_{pe}^2}  
\end{equation} 
where ${\rm E} _{kin} $ is the density of the ion kinetic energy,  $W_{0} =|E_{0} |^{2} /4\pi $  is the initial energy density of long-wavelength Langmuir waves \cite{Belkin.2013}.

The ratio of characteristic time scales for these two models is of the order of $(m_{e} /M)^{1/6} (W/n_{0} T_{e} )^{1/2}$. Considering this, it was found that the kinetic energy of ions to be of the same order in both models. The ratio of ion kinetic energy to the initial energy of long-wave oscillations occurs equal to ${\rm E} _{kin} /W_{0} \sim \delta /\omega _{pe} $ for Silin's model and ${\rm E} _{kin} /W_{0} \simeq 0.5\cdot W_{0} /n_{0} T_{e} $ for Zakharov's model. This means that in the Silin model, the ions acquire a portion of field energy of the order of $\delta /\omega _{pe}$. This effect was predicted in \cite{Andreev.1977} and confirmed in \cite{Kuklin.1990}. A portion of transferred energy in Zakharov's model is of the order of $W_{0}/n_{0} T_{e}$.


\begin{figure}[p]\centering
\subfigure[Silin model]{\includegraphics[width=0.47\textwidth]{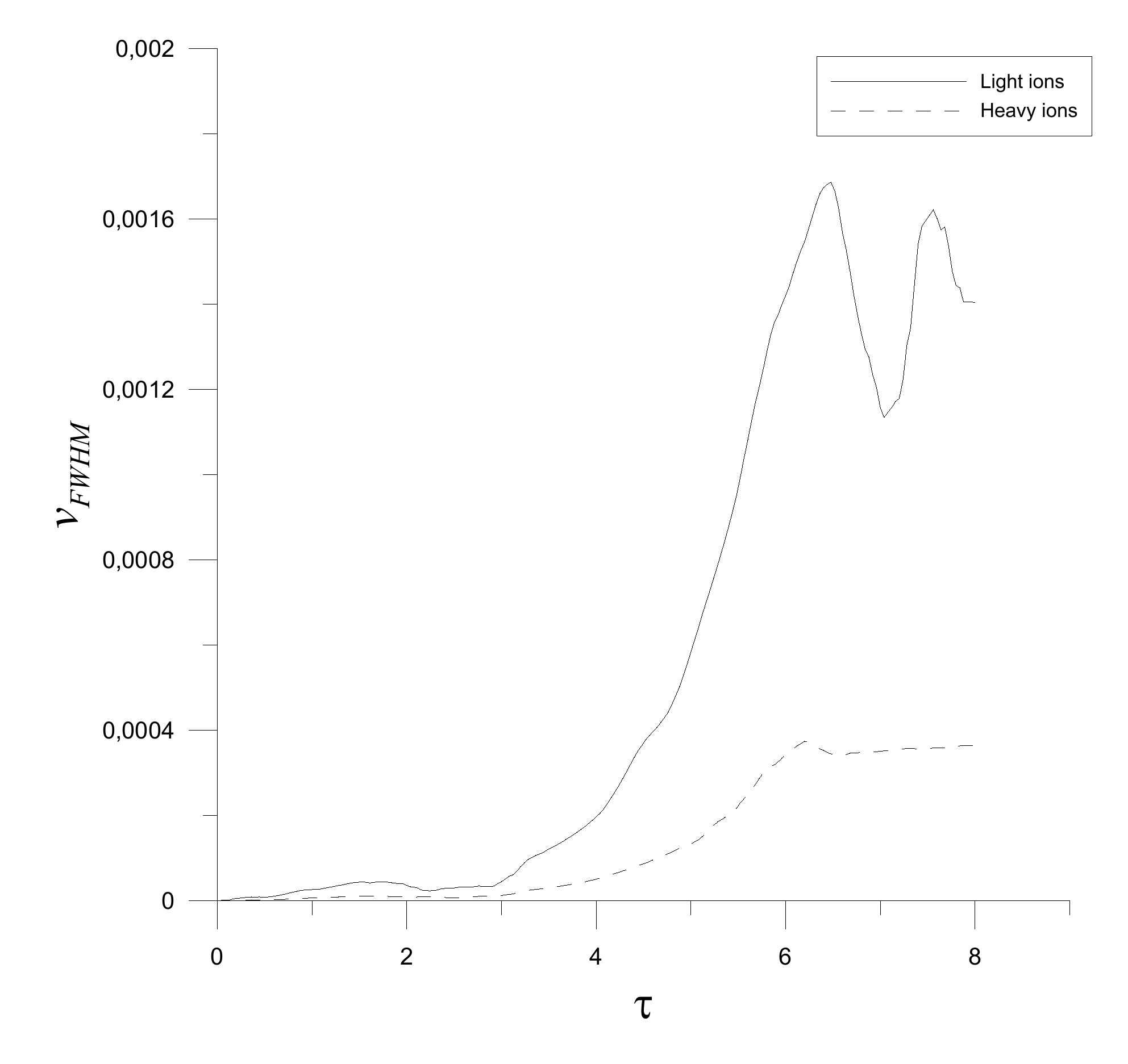}}  \hfill
\subfigure[Zakharov model]{\includegraphics[width=0.47\textwidth]{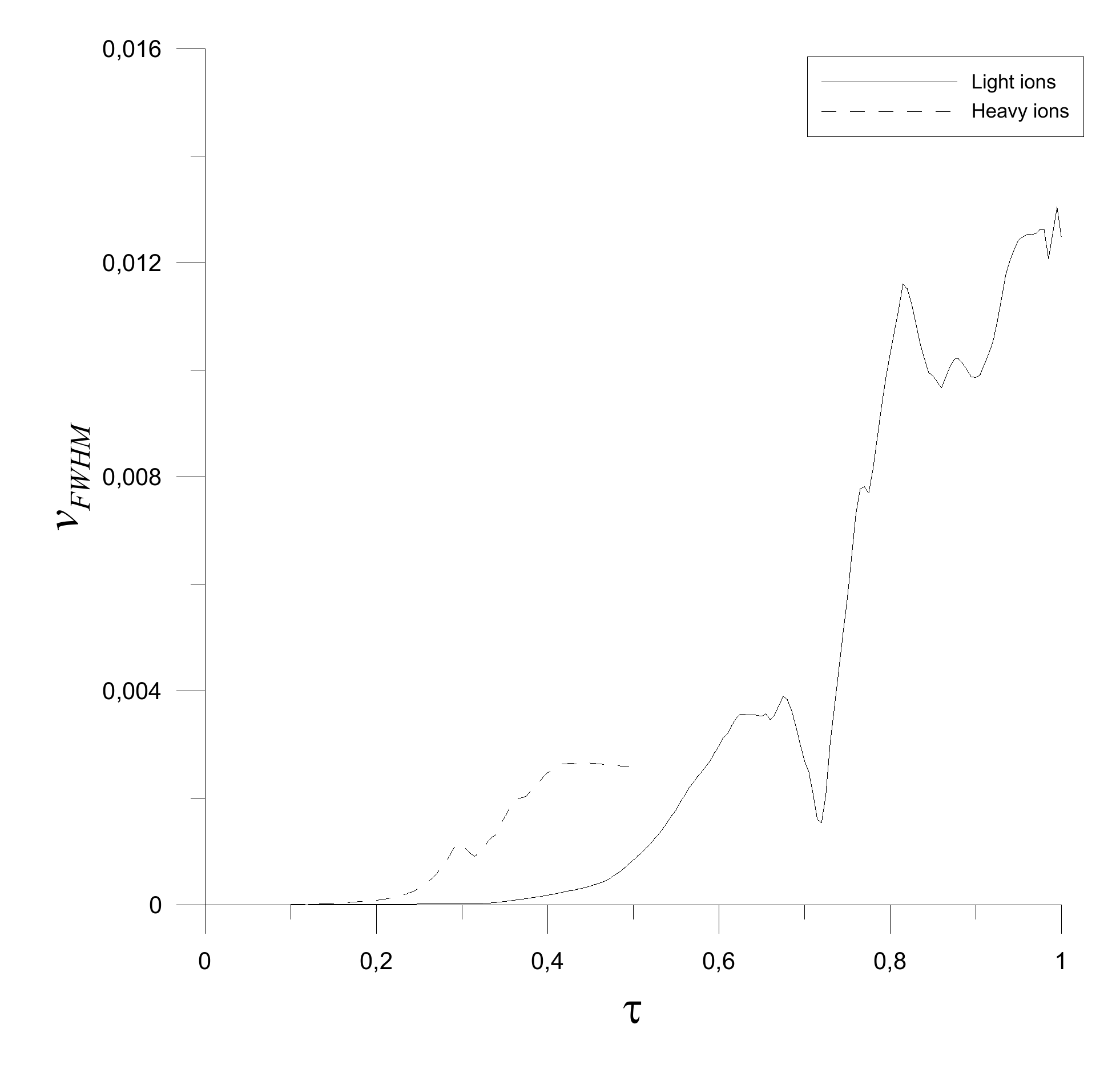}} \\

\caption{Evolution of $v_{\textrm {\tiny FWHM}}$ (full width at half maximum) of the ion velocity distribution function $f(v_s)$.}
 \label{fig3}
\end{figure}  

\begin{figure}[p]\centering
\subfigure[Silin model]{\includegraphics[width=0.47\textwidth]{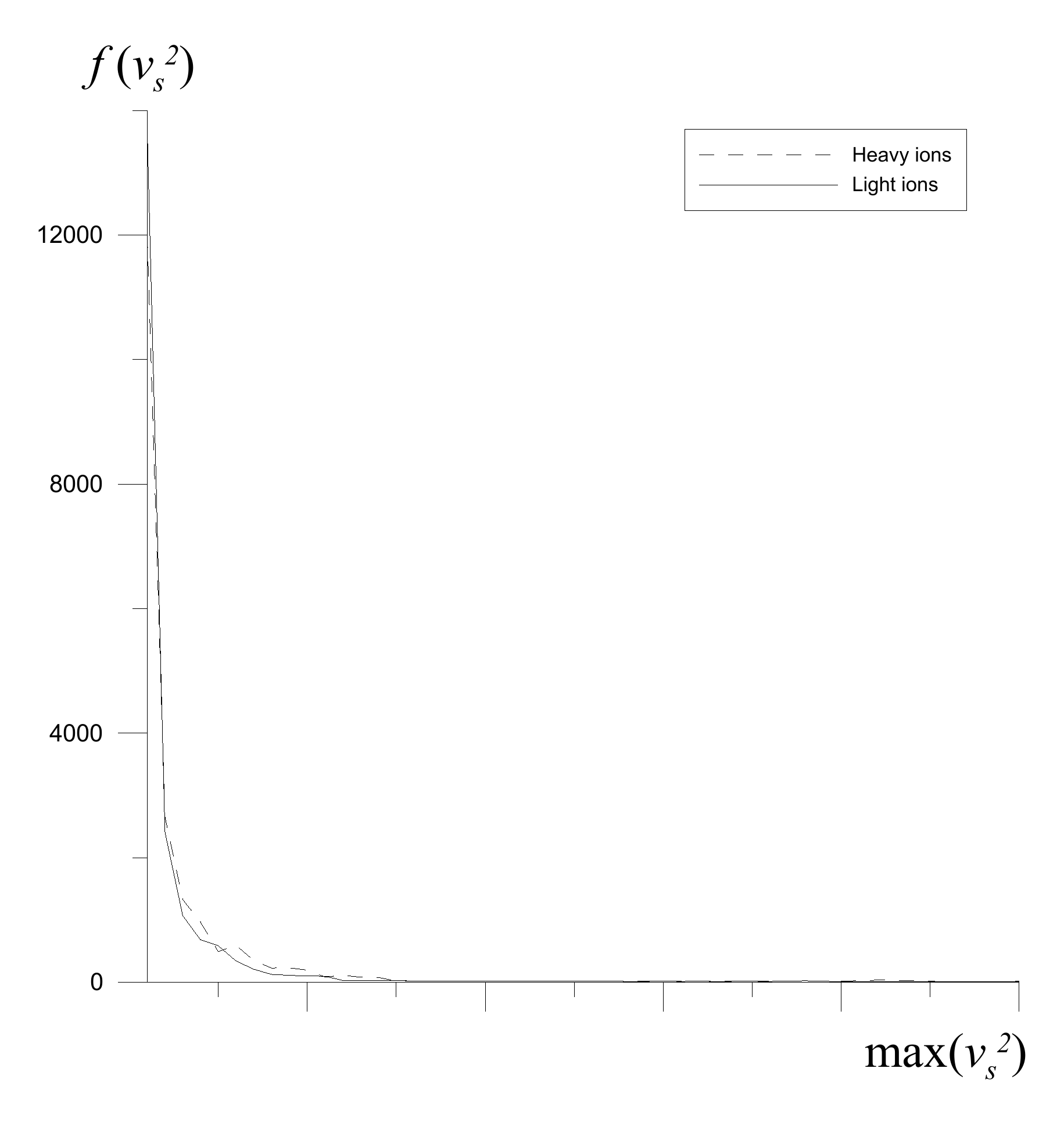}}  \hfill
\subfigure[Zakharov model]{\includegraphics[width=0.47\textwidth]{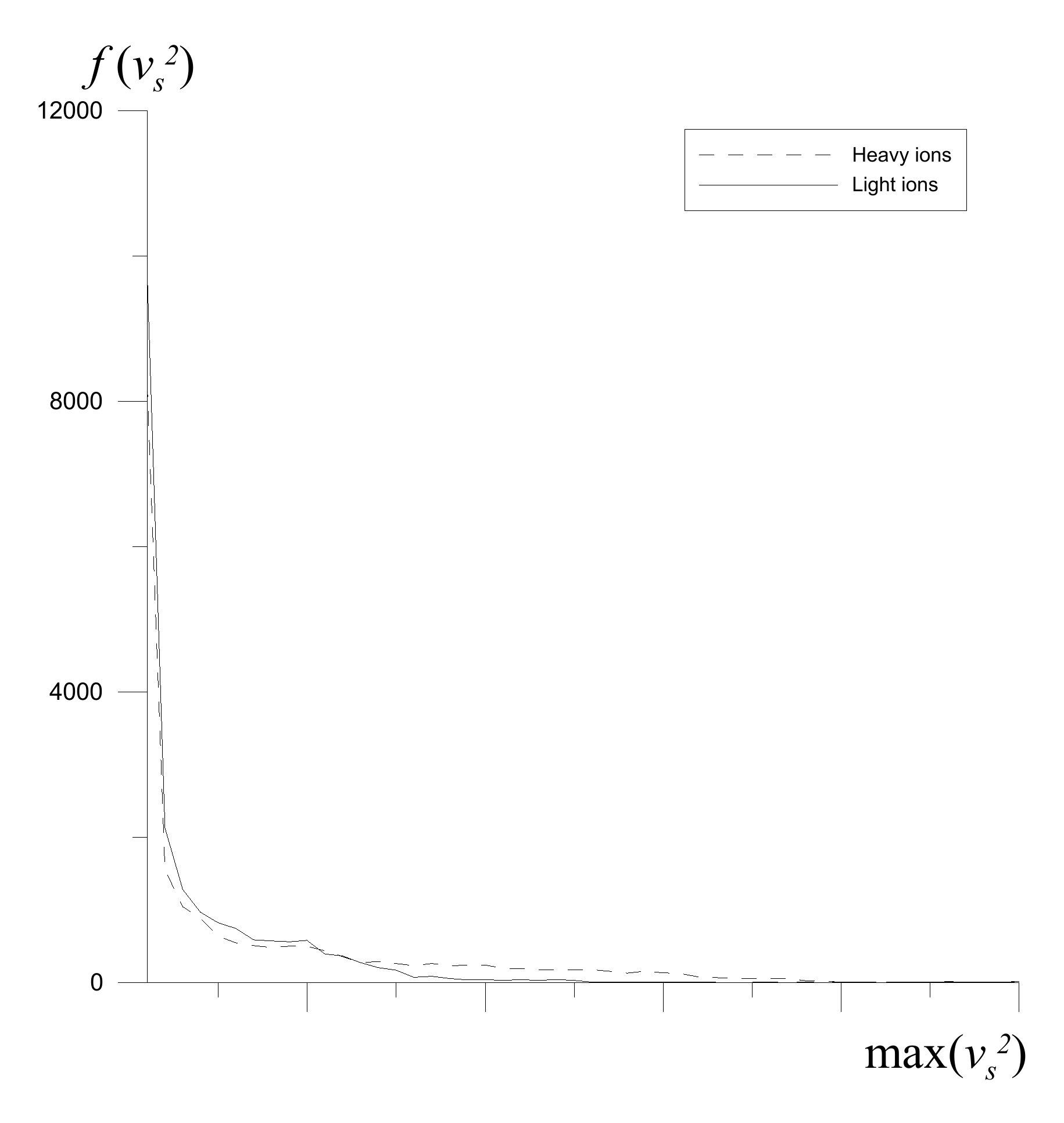}} \\

\caption{Distribution function of ion super-particles over $v_s^2=(d\xi _{s} /d\tau )^{2}$.}
 \label{fig4}
\end{figure}

\section{Analysis of ion velocity distribution}

Let consider approximation of the ion velocity distribution function obtained from numerical experiments by the normal distribution function.  Generally speaking, There three ways to fit a normal (Gaussian) distribution (Gaussian distribution) to the simulated ion velocity distribution.

\begin{figure}[h]\centering
\subfigure[Ion velocity distribution, obtained from numerical experiment]{\includegraphics[width=0.47\textwidth]{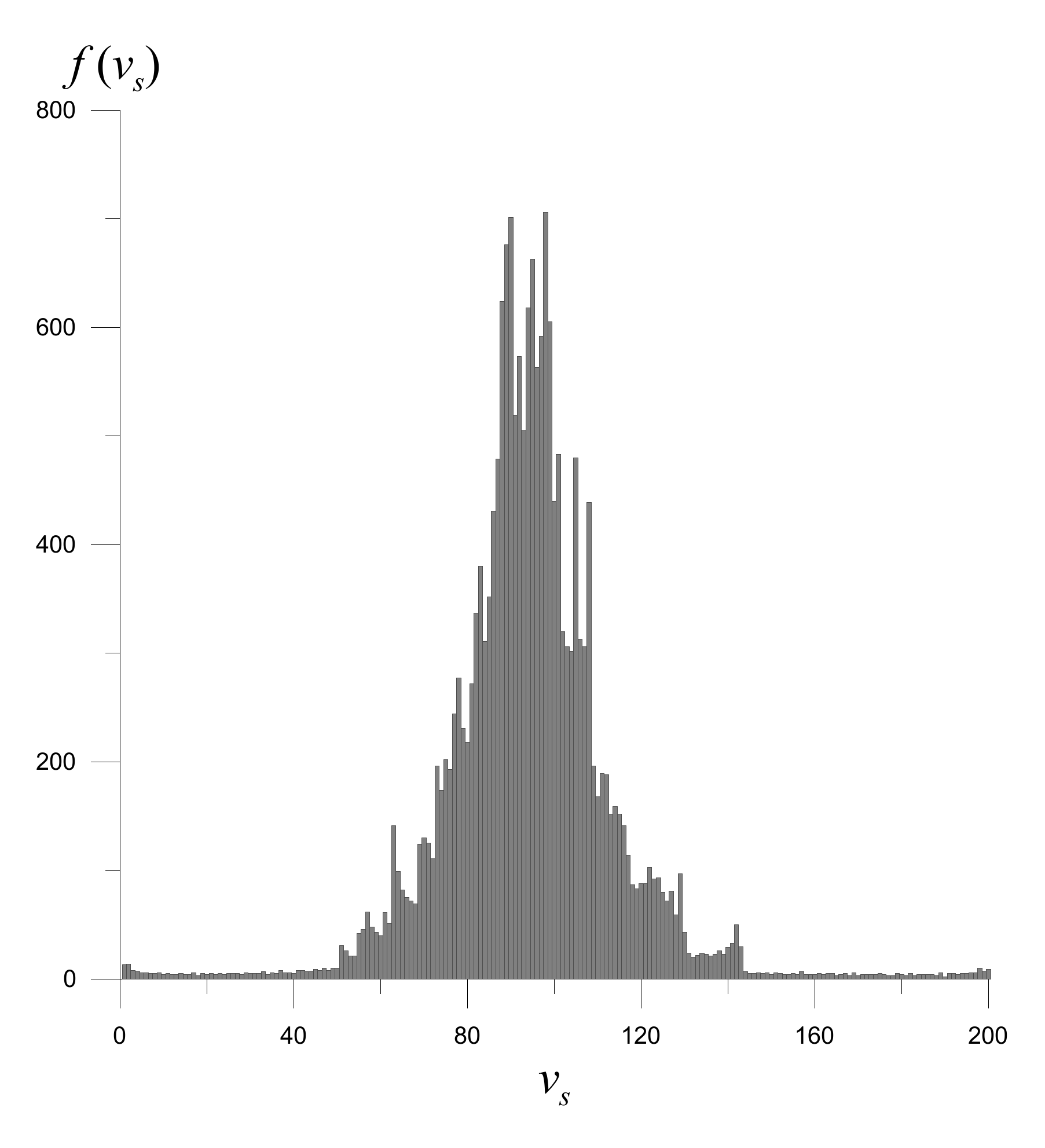}}  \hfill
\subfigure[Normal ion velocity distribution]{\includegraphics[width=0.47\textwidth]{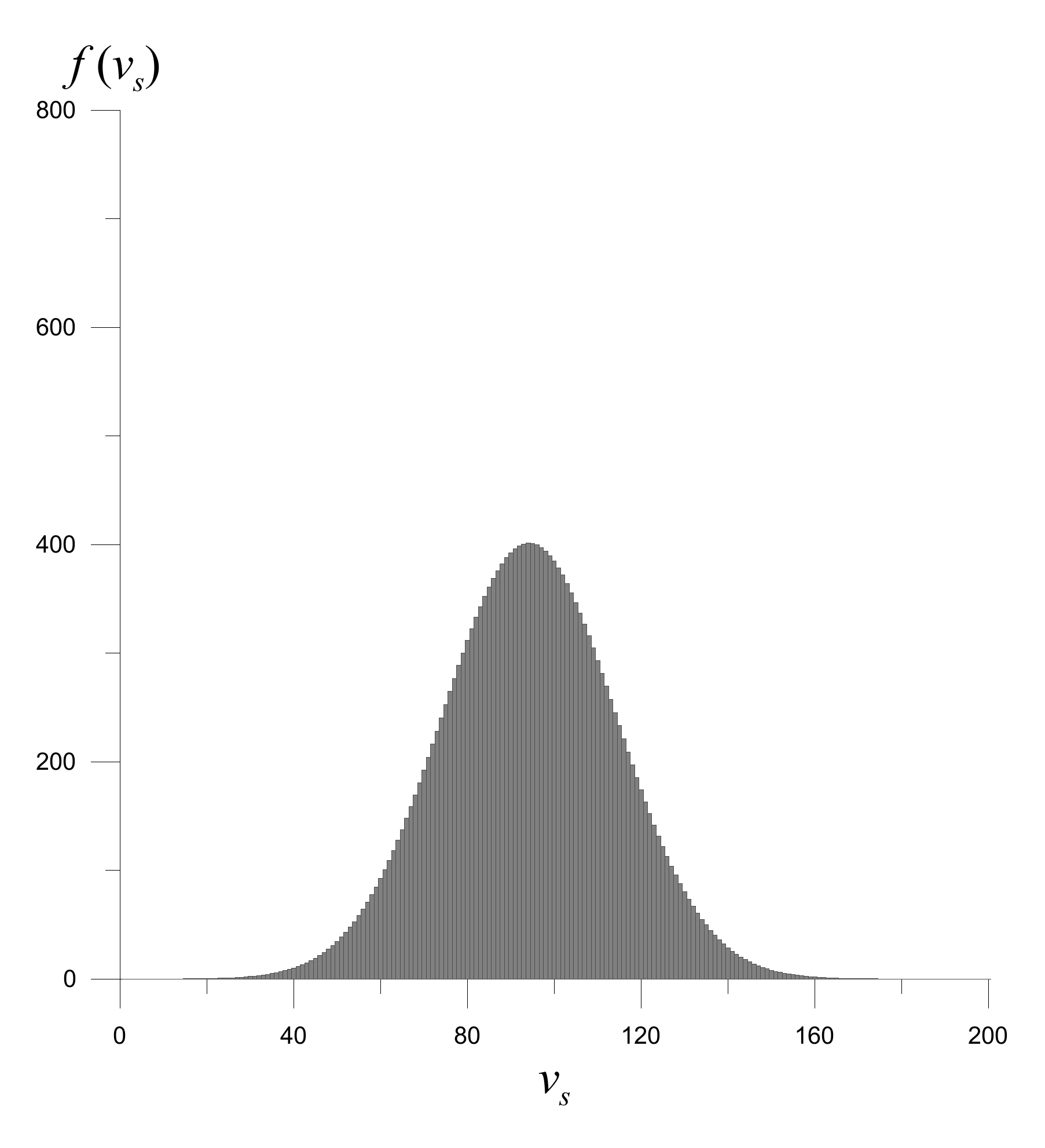}} \\

\caption{A sample of ion velocity distribution histogram and corresponding normal statistical distribution that best fits to it.}
 \label{fig5}
\end{figure}

\begin{figure}[h]\centering
\subfigure[$N_1^{(out)}$ -- the number of particles which belong to the calculated distribution and are outside of the normal distribution]{\includegraphics[width=0.47\textwidth]{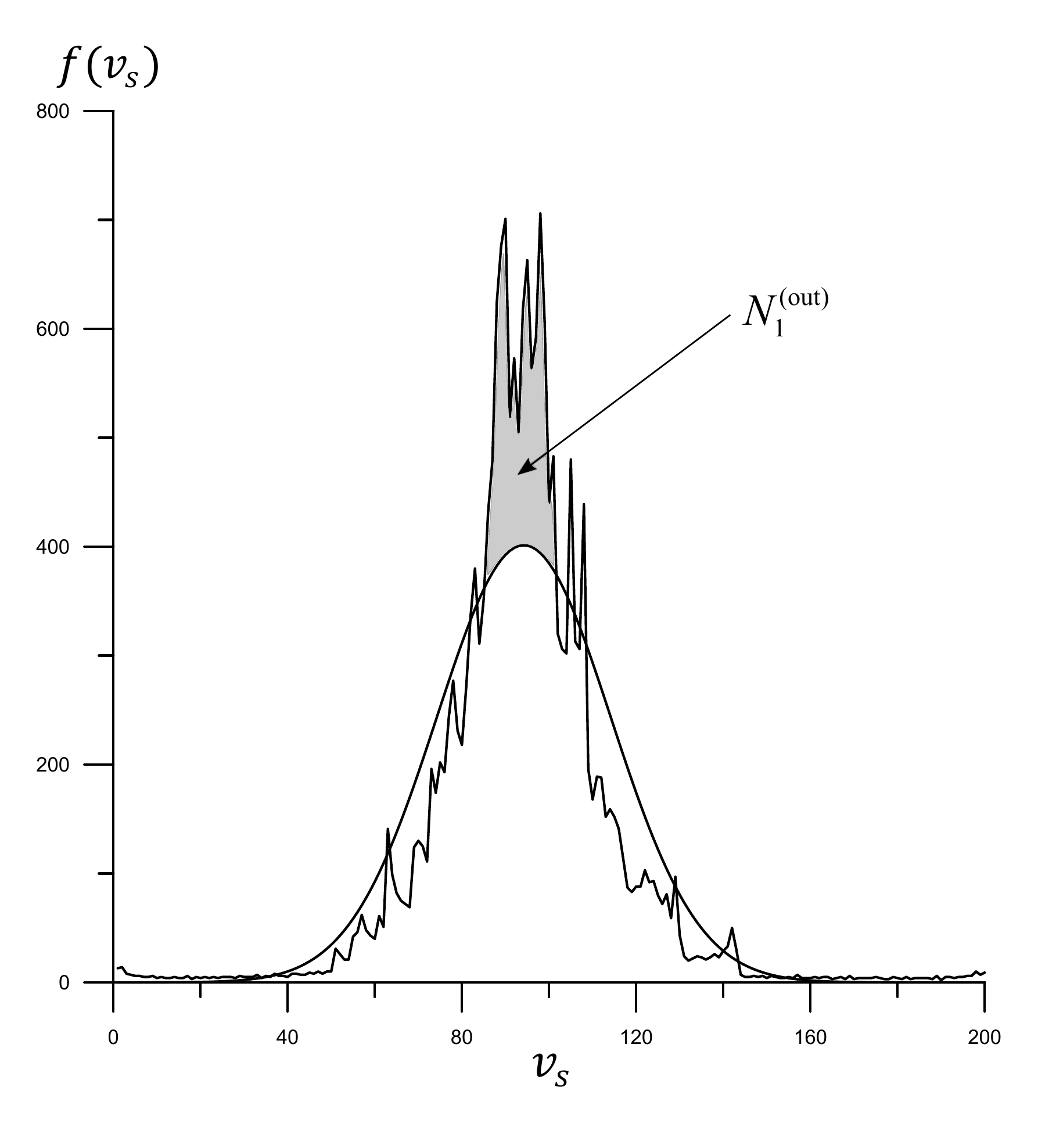}}  \hfill
\subfigure[$N_2^{(out)}$ -- the number of particles which belong to the normal distribution and are outside of the calculated distribution]{\includegraphics[width=0.47\textwidth]{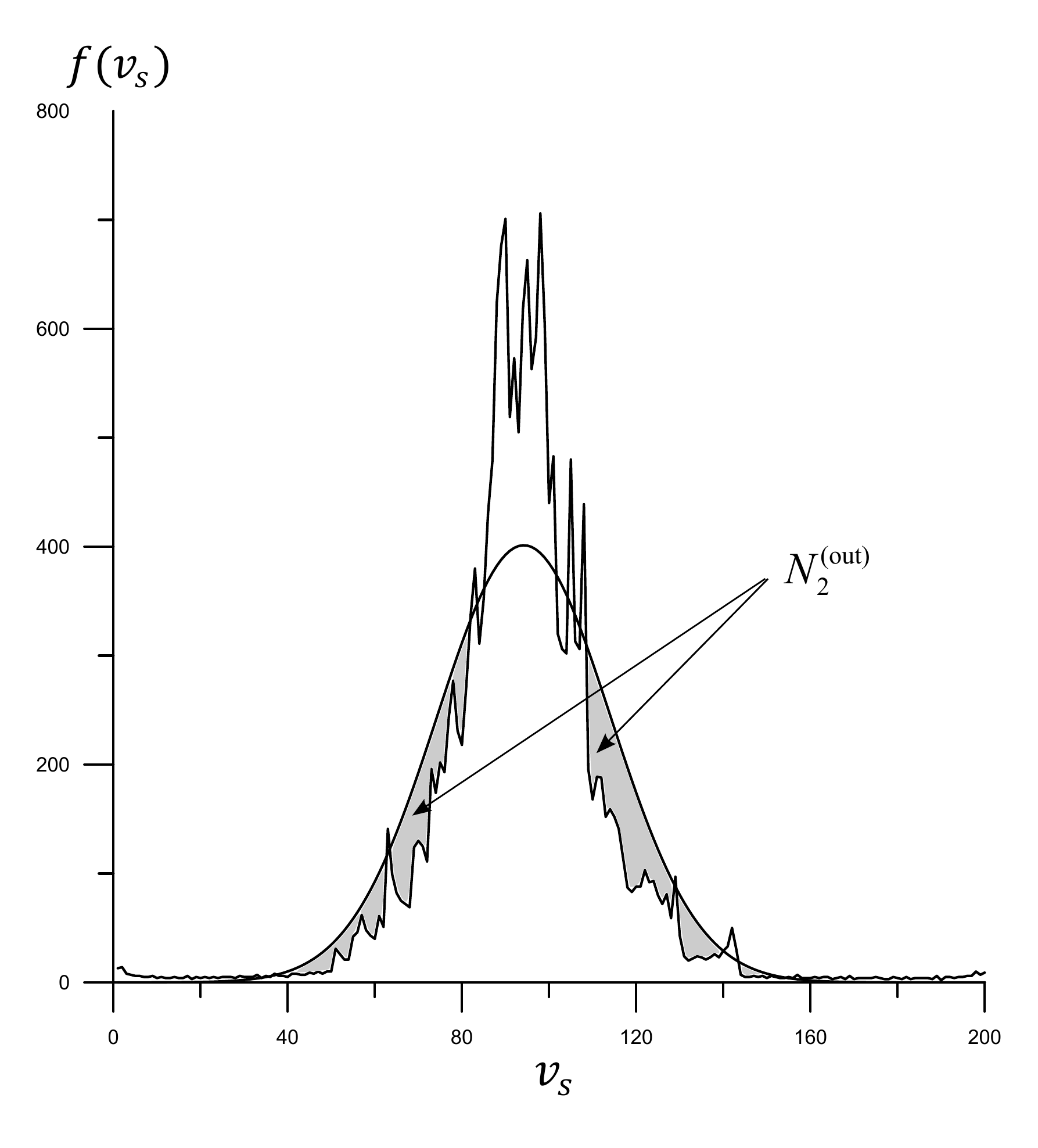}} \\

\caption{Illustration to the explanation of methods of distribution fitting}
 \label{fig6}
\end{figure}

A normal distribution density function is defined as

\begin{equation} \label{eq15} 
f(x)=\frac{1}{\sigma \sqrt{2\pi } } e^{-\frac{(x-\mu )^{2} }{2\sigma ^{2} } } , 
\end{equation} 

\noindent where $\mu ={\sum v_{s}}/{S} $ is the mean or expectation, $\sigma ^{2} $ is the variance, $\sigma $ is standard variation, $x$ is the ion velocity, which varies from minimum to maximum value in steps of 1/200. 

In order to plot the histogram we calculate the number of ions in a point $x$ by the formula $$S_{x} =\frac{f(x_i)} {\sum _{i=1}^{200}f(x_i)}\cdot S$$ and plot it as ordinate.

The first method of distribution fitting by Gaussian distribution function implies calculation of the standard deviation of the particle velocities $v_{s}$:

\begin{equation} \label{eq16} 
\sigma =\sqrt{\frac{1}{S} \sum \left(v_{s} -\mu \right)^{2}  } .      
\end{equation}

The second method of normal distribution fitting is based on choosing of standard deviation $\sigma$ as $0.9 v_{\textrm {\tiny FWHM}}$ ( $v_{\textrm {\tiny FWHM}}$ is the full width at half maximum, see Fig.\ref{fig3}).  

The third method is similar to the second one, but $\sigma = k v_{\textrm {\tiny FWHM}}$, where   coefficient $k$ is chosen in such a way that the difference between simulated and normal distribution will be minimal. The difference between these two distributions is defined as $(N_1^{(out)}+N_2^{(out)})/N* 100\%$ (see Fig.\ref{fig6}).

\begin{figure}[h]\centering
\subfigure[Silin model]{\includegraphics[width=0.47\textwidth]{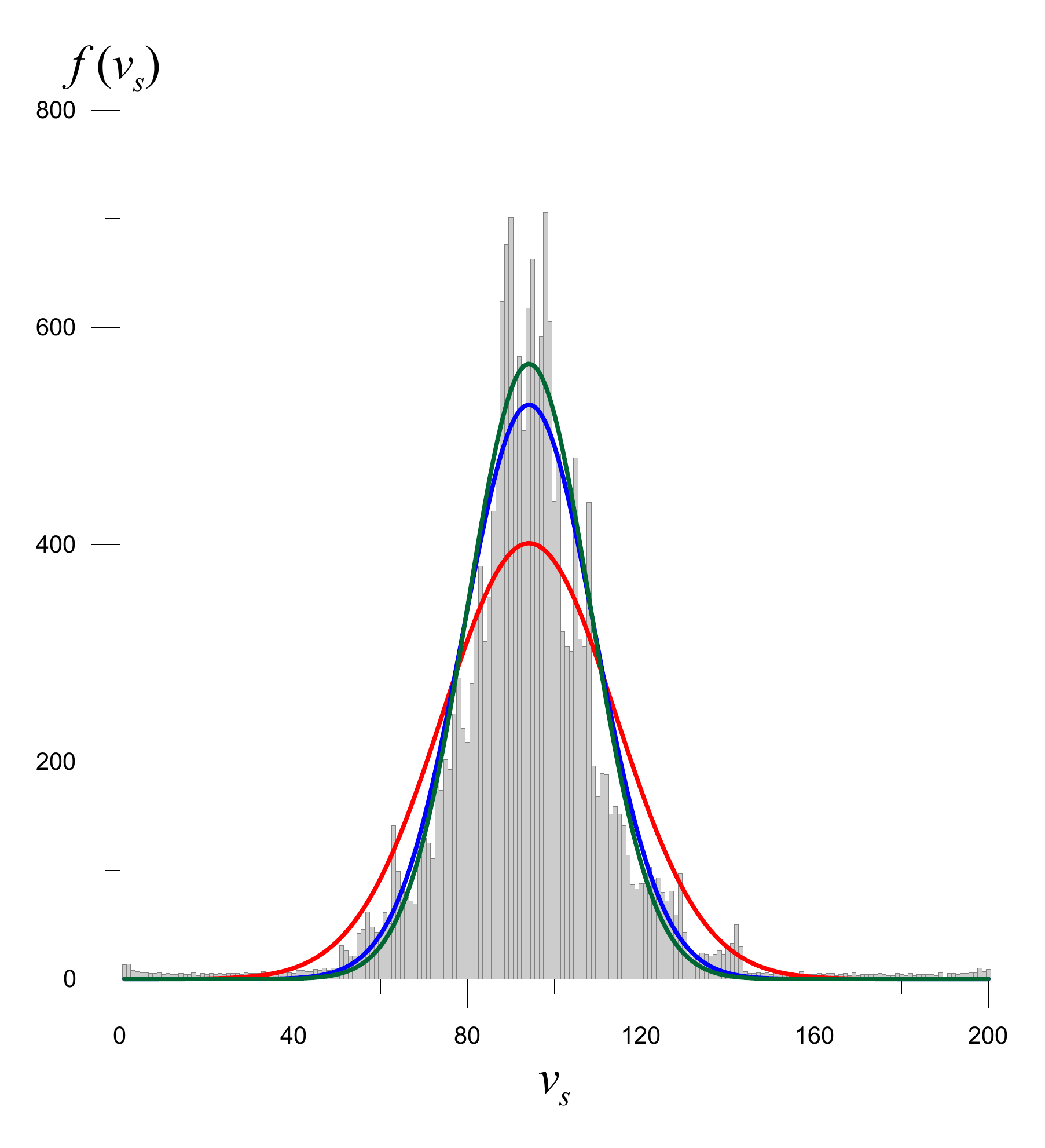}}  \hfill
\subfigure[Zakharov model]{\includegraphics[width=0.47\textwidth]{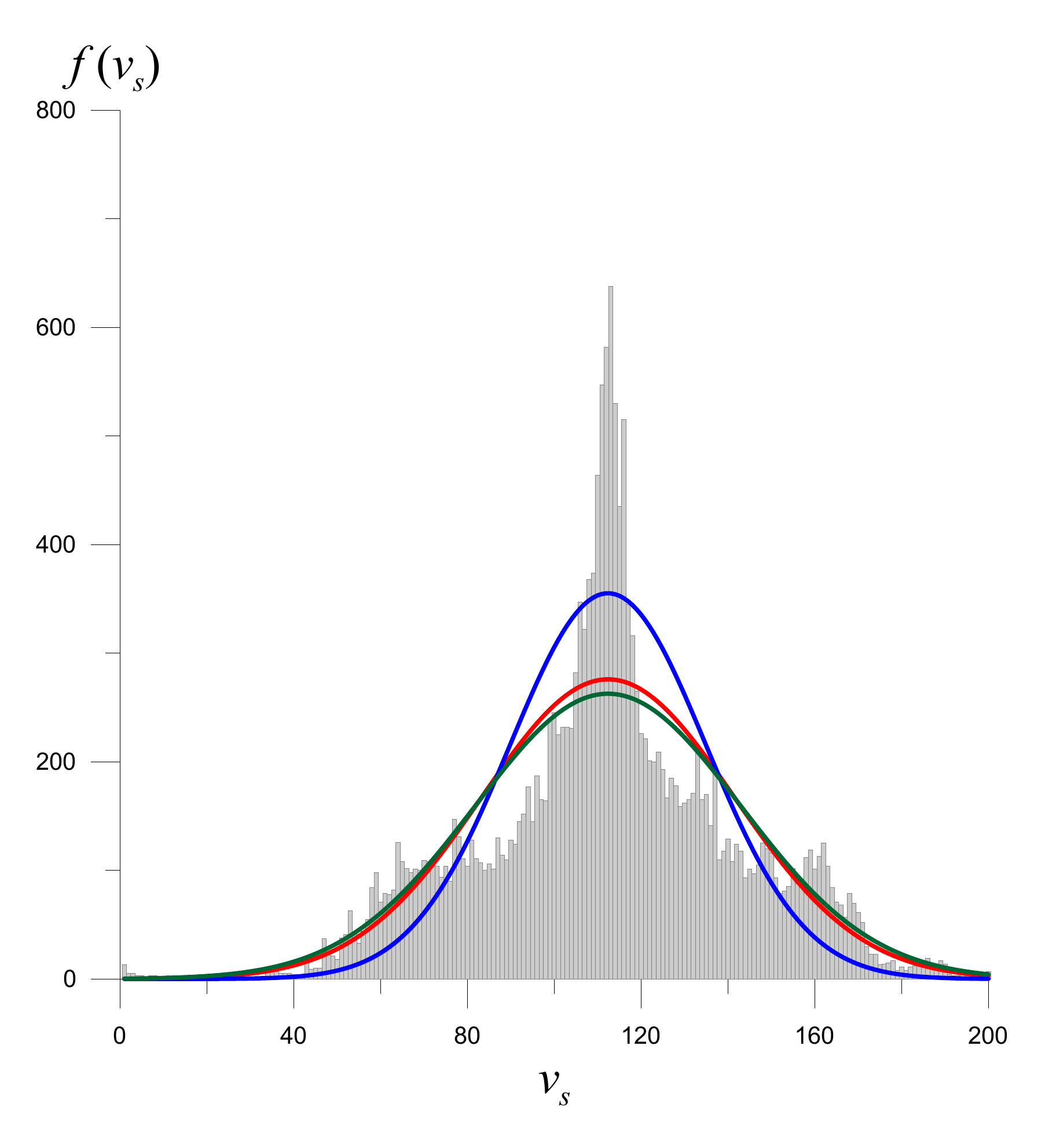}} \\

\caption{Normal distribution fitting by three methods: red curve -- the first method, blue curve -- the second method, green curve -- the third method}
 \label{fig7}
\end{figure}

\subsection{Ion velocity distribution for the Silin hybrid model}
 
The simulation of ion kinetics within framework of the Silin model for the case of light ions gives the results shown in the Fig.\ref{fig7}a. 
 


It is also interesting to find out the effect of different distributions of initial RF wave amplitudes and phases on the parameters of final ion velocity distribution. Table \ref{tab2} shows that initial conditions do not lead to a qualitative change in the ion velocity distribution.

\begin{table}[t] \footnotesize
\caption{Results of normal distribution fitting for the Silin model}
\newcolumntype{R}{>{\raggedleft\arraybackslash}X}%
\begin{tabularx}{\textwidth}{p{5cm}RRRRRR} \hline \hline 
 & \multicolumn{3}{c}{Light ions } & \multicolumn{3}{c}{Heavy ions} \\ 
 & method 1 & method 2 & method 3 & method 1 & method 2 & method 3 \\ \hline
The difference between the normal curve and calculated distribution & 18.4\% & 12.9\% & 12.7\% & 17.4\% & 15.1\% & 14.2\% \\ \hline 
Coefficient $k$ \par (from $\sigma =k \max (\bar{v})$) & - & 0.9 & 0.84 & - & 0.9 & 1.07 \\ \hline 
Total kinetic energy of particles outside of the normal distribution $E_{kin}^{(out)}$& 0.0287 & 0.0421 & 0.0463 & 0.0017 & 0.0032 & 0.0025 \\ \hline 
Total kinetic energy of particles that are absent inside the normal distribution $E_{kin}^{(in)}$ & 0.0273 & 0.0073 & 0.0057 & 0.0016 & 0.0003 & 0.0005 \\ \hline 
$E_{kin}^{(out)}-E_{kin}^{(in)}$ & 0.0014 & 0.0348 & 0.0406 & 0.0001 & 0.0029 & 0.002 \\ \hline 
Total kinetic energy of particles $E_{kin}^{(total)}$& \multicolumn{3}{c}{0.08} & \multicolumn{3}{c}{0.005} \\ \hline 
$(E_{kin}^{(out)}-E_{kin}^{(in)})/E_{kin}^{(total)}$ & 0.02 & 0.44 & 0.51 & 0.02 & 0.58 & 0.4 \\ \hline 
$E_{kin}^{(in)}/E_{kin}^{(total)}$ & 0.36 & 0.53 & 0.58 & 0.34 & 0.64 & 0.5 \\ \hline 
\end{tabularx}
\label{tab2} 
\end{table}

\subsection{Ion velocity distribution for the Zakharov hybrid model} 

The simulation of ion kinetics within framework of the Zakharov model gives the following distribution, at the moment when the growth of ion energy is stopped, shown in the Fig.\ref{fig7}b.



 As clear from the Table \ref{Tab3} the initial conditions of initial RF wave amplitudes and phases do not lead to qualitative change in the ion velocity distribution. 

{\noindent\small
\begin{table}[t]  \footnotesize
\caption{Results of normal distribution fitting for the Zakharov model}
\newcolumntype{R}{>{\raggedleft\arraybackslash}X}%
\begin{tabularx}{\textwidth}{p{5cm}RRRRRR} \hline \hline 
& \multicolumn{3}{c}{Light ions} & \multicolumn{3}{c}{Heavy ions} \\
& method 1 & method 2 & method 3 & method 1 & method 2 & method 3 \\ \hline \noalign{\smallskip} 
The difference between the normal curve and calculated distribution & 16.2\% & 20.2\% & 15.9\% & 17.5\% & 37.6\% & 16.9\% \\ \hline 
Coefficient $k$ \par (from $\sigma =k \max (\bar{v})$) & - & 0.9 & 1.22 & - & 0.9 & 2.67 \\ \hline \noalign{\smallskip}
Total kinetic energy of particles outside of the normal distribution $E_{kin}^{(out)}$ & 0.866 & 2.27 & 0.655 & 0.178 & 0.745 & 0.084 \\ \hline \noalign{\smallskip}
Total kinetic energy of particles that are absent inside the normal distribution $E_{kin}^{(in)}$ & 0.741 & 0. 386 & 0.966 & 0.104 & 0.032 & 0.151 \\ \hline \noalign{\smallskip}
$E_{kin}^{(out)}-E_{kin}^{(in)}$ & 0.125 & 1.884 & 0.311 & 0.074 & 0.713 & 0.067 \\ \hline 
Total kinetic energy of particles $E_{kin}^{(total)}$& \multicolumn{3}{c}{4.584} & \multicolumn{3}{c}{0.831} \\ \hline \noalign{\smallskip}
$(E_{kin}^{(out)}-E_{kin}^{(in)})/E_{kin}^{(total)}$ & 0.03 & 0.41 & 0.07 & 0.09 & 0.86 & 0.08 \\ \hline \noalign{\smallskip}
$E_{kin}^{(in)}/E_{kin}^{(total)}$ & 0.19 & 0.5 & 0.14 & 0.21 & 0.9 & 0.1 \\ \hline\hline 
\end{tabularx}
\label{Tab3}
\end{table}
}

For the Maxwellian velocity distribution the half-width (see Fig.\ref{fig5}) of the distribution will be related with thermal velocity by the relation $\bar{v}=1.18 v_{T}$. As follows from calculations the root-mean-square velocity measured in relation units after the saturation of ions energy is equal $\sqrt{<v_{s}^{2} >} =\sqrt{I_{s}/S} $. If this value is of the order of  $0.85 \bar{v}$, e.g. $\sqrt{I_{s} /S} \approx 0.85 \bar{v}$, than it is reasonably to consider the distribution as Maxwellian and one can say about the ion temperature. When $\sqrt{I_{s} /S} >0.85 \bar{v}$, the distribution has a "tail" of fast particles. It follows from simulation results that in the hot plasma (the Zakharov model) the ion velocity distribution is close to Maxwellian and one can say about the ion temperature $T_{i} \sim W_{0}^{2} /n_{0}^{2} T_{e} $. In the case of cold plasma the ion velocity distribution contains noticeable fraction of fast particles, that was observed in experiments \cite{Batanov.1986}.


We can fit parameters of the normal distribution in such a way that the number of particles outside this distribution will be minimal. We can also define the ratio of total kinetic energy of particles outside the normal distribution to the total kinetic energi of modeling particles (see Table \ref{Tab4}).

\noindent
\begin{table}[t] \footnotesize
\caption{Calculation of ion kinetic energy}
\begin{tabularx}{\textwidth}{p{5cm}XXXX} \hline \hline 
          Characteristics  & \multicolumn{2}{c}{Hot plasma} & \multicolumn{2}{c}{Cold plasma} \\ 
            & Light ions & Heavy ions & Ligt ions & Heavy ions\\  \hline \noalign{\smallskip}
$I=\sum _{s}\left({d\xi _{s} }/{d\tau } \right) ^{2} $ & 4.58 & 0.831   & 0.0803 & 0.00512 \\ \noalign{\smallskip}\hline \noalign{\smallskip}
The sum of squared velocities $d\xi _{s} /d\tau $ of particles outside of the normal distribution  & 0.655   & 0.084  & 0.0463 & 0.00251 \\\noalign{\smallskip} \hline \noalign{\smallskip}
The sum of squared velocities  $d\xi _{s} /d\tau $ of particles that are absent inside the normal distribution & 0.966  & 0.151  & 0.00567 & $5,4 10^{-4} $ \\ \noalign{\smallskip}\hline \noalign{\smallskip}
The difference between the normal curve and calculated distribution &  15,9\%  &  16,9\% &  12,7\% &  14,2\% \\ \noalign{\smallskip}\hline \noalign{\smallskip}
The number of particles outside the normal distribution &  3178 &  3386 &  2549 &  2846 \\ \noalign{\smallskip} \hline \noalign{\smallskip} 
The ratio of total kinetic energy of ions to initial field energy ${\rm E} _{kin} /W_{0} $ & $3,1 10^{-2} $  & $0,57 10^{-2} $ & $5 10^{-2} $ & $1,34 10^{-2} $ \\\noalign{\smallskip} \hline 
\end{tabularx}
 \label{Tab4}
\end{table}

For hot plasma (Zakharov's model) this ratio is of the order of 10-14\% and for cold plasma this ratio can be estimated as 50-60\%. Thus, the ion kinetic energy distribution is close to maxwellian in the case of hot plasma and in the case of cold plasma the fast ions are approximately of a half of total kinetic energy of ions.

 Note in conclusion that perturbations of ion density with scales less than the ion Debye radius $r_{Di} =v_{Ti} /\omega _{pi} $ don't contribute to formation of low-frequency electric fields due to the screening effect. 
 In terms of $R_{Di}=r_{Di} k_{0} /2\pi $ the ion Debye radius can be evaluated as 

\begin{equation} \label{eq20}
R_{Di} \sim \left<\frac{v_{i} k_{0} }{2\pi \gamma _{L} } \right>\left(\frac{\delta }{\omega _{pe} } \right)\left(\frac{M}{m_{e} } \right)^{1/2} =\left<v_{s} \right>\left(\frac{\delta }{\omega_{pe} } \right)\left(\frac{M}{m_{e} } \right)^{1/2}.  
\end{equation}

On the stage of the developed instability this value occurs of the order of $R_{Di} \le 10^{-3} $, and the number of modes in the spectrum of ion density doesn't exceed $1/R_{Di} $, which is consistent with our analysis.

\section{Discussion}

The mechanisms of modulation instability of long-wave Langmuir oscillations both in hot and in cold plasma have much in common. The spectra of growing perturbations have the same symmetry \cite{Silin.1965}, \cite{Kuklin.2013}, the mechanisms of their growth are similar too.
The energy transfer through the spectrum in the Zakharov and Silin models is caused not only by the rearrangement of the field due to interaction between spectrum modes, but it is largely a result of the linear instability. Maximum growth rate in the Zakharov model increases with decreasing of perturbation scale. In the Silin model, the maximum growth rate shifts to the short-wavelength area with decreasing of the pump wave amplitude \cite{Silin.1965} that is confirmed by results of the nonlinear theory. It is important that the maximum growth rate remains unchanged with decreasing of the pump wave amplitude in the case of cold plasma and decreases in whole instability domain in the case of hot plasma. The Langmuir oscillations of large amplitude, excited by high-current beam of charged particles, have a wavelength that exceeds a maximum wavelength of the excited spectrum by no more than two orders. Therefore, there is no practical sense to consider smaller-scale perturbations. 

The most important consequence of the development of parametric instabilities of intense Langmuir waves in plasma is the energy transfer from the electric field to plasma ions (ion heating). It is reasonably to consider this problem within the framework of hybrid models, where the electrons are described as fluid and ions - kinetically, i.e. as super particles. Analysis of such hybrid models shows that in the case of hot plasma the ions get a portion of field energy that is proportional to the ratio of the field energy to the plasma thermal energy. In the case of cold plasma the ions get a portion of field energy that is proportional to ratio of the instability growth rate to the plasma frequency or that the same to the cubic root of ratio of electron mass to ion mass. The energy transferred to ions in the case of heavy ions is significantly less than in the case of light ions. 
 	 

It is also shown that the kinetic energy distribution of ions in the Zaharov hybrid model is close to Maxwellian and we can talk about the temperature of the ions.
The kinetic energy distribution of ions in the Silin hybrid model differs substantially from Maxwellian and is characterized by a large fraction of fast particles.
  
This paper was partially supported by the grant of the State Fund for Fundamental Research (project No. \selectlanguage{russian}Ф\selectlanguage{english}58/175-2014). The authors thank Prof.~V.I.~Karas’ for helpful comments.

\section*{References}
\bibliographystyle{elsarticle-num} 
\bibliography{pryjmak4} 

\end{document}